\documentclass[12pt]{iopart}
\usepackage{iopams}
\usepackage{epsf}
\def\r{\mathbb R}                   % Numeros reales
\def\NAT{\mathbb N}                   % Numeros naturales
\def\d{\partial}
\def\fr{\frac}
\def\be{\begin{equation}}
\def\ee{\end{equation}}
\def\bea{\begin{eqnarray}}
\def\eea{\end{eqnarray}}
\def\b*{\begin{eqnarray*}}
\def\e*{\end{eqnarray*}}

\def\T{{\bf T}}
\def\S{{\bf S}}

\def\Q{{\bf Q}}
\def\G{{\bf g}}
\def\R{{\bf R}}

\def\DP{{\cal DP}}

\def\C{{\cal C}}
\def\b{\bar}
\def\g{\gamma}

\def\O{{\bf\Omega}}
\def\f{\varphi}

\def\k{\vec{k}}

\def\xiv{\vec \xi}

\def\lie{{\pounds}_{\xiv}\, }

\def\u{\vec{u}}

\def\K{{\bf k}}

\newtheorem{lem}{Lemma}

\newtheorem{res}{Result}
\begin{document}
\title[Causal symmetries]{Causal symmetries}

\author{Alfonso Garc\'{\i}a-Parrado and Jos\'{e} M M Senovilla}

\address{F\'{\i}sica Te\'{o}rica, 
Universidad del Pa\'{\i}s Vasco, Apartado 644, 48080 Bilbao, Spain}

\eads{\mailto{wtbgagoa@lg.ehu.es} and \mailto{wtpmasej@lg.ehu.es}}

\begin{abstract}
Based on the recent work \cite{PII} we put forward
a new type of transformation for Lorentzian manifolds
characterized by mapping every causal future-directed vector onto a
causal future-directed vector. The set of all such transformations,
which we call causal symmetries, has the structure of a submonoid
which contains as its maximal subgroup the set of conformal
transformations. We find the necessary and sufficient conditions
for a vector field $\xiv$ to be the infinitesimal generator
of a one-parameter submonoid of pure causal symmetries. We speculate about
possible applications to gravitation theory by means of some
relevant examples.
\end{abstract} 
Our goal is to introduce a new type of spacetime symmetry 
which generalizes the conformal one while still
preserving many causal properties of the
Lorentzian manifolds. To that end, we will need the results on
null-cone preserving maps analyzed and classified in \cite{PI}.
The whole idea will be based on the new concept of {\em causal
mapping} (leading to a definition of
{\em isocausal spacetimes}) which was recently introduced in \cite{PII}.
This letter is inspired by \cite{PI} and \cite{PII}
which will be referred to as PI and PII from now on, respectively,
and we use their notations. Herein, we will just give the
fundamental results. A longer detailed exposition
will be given elsewhere \cite{S}. Some related ideas
were used in \cite{LOW}.

According to PII, a causal relation between two Lorentzian manifolds
is any diffeomorphism which maps non-spacelike (also called causal)
future-directed vectors onto causal future-directed vectors. Here we 
will say that a transformation
$\f:(V,\G)\rightarrow (V,\G)$ is a {\em causal symmetry} if it sets a
causal relation of $(V,\G)$ with itself. From theorem 3.1 in PII
follows that $\f$ is a causal symmetry iff $\f^{*}\G$ satisfies the
dominant energy condition, or in the notation of PI and PII,
iff $\f^{*}\G$ is a future tensor: $\f^{*}\G\in\DP^{+}_2(V)$.

The set of causal symmetries of $(V,\G)$ will be denoted by $\C(V,\G)$
(in short $\C(V)$ if no confusion arises).  This is a subset of the 
transformation group of $V$ and clearly (prop.\ 3.3 of PII)
the composition of causal symmetries is a causal symmetry. As
the identity map is also a causal symmetry, $\C(V)$ has the
algebraic structure of a submonoid, see e.g. \cite{SEMIGROUP}.
Nonetheless $C(V)$ will not in general be a subgroup because
the inverse of a causal relation need not be a causal relation.
Actually, both $\f$ and $\f^{-1}$ are causal iff 
$\f$ is a conformal transformation (theorem 4.2 of PII), and
therefore the maximal subgroup $\C(V)\cap\C(V)^{-1}$ of $\C(V)$ \cite{SEMIGROUP}
is just the group of conformal transformations of $V$: 
every subgroup of $\C(V)$ is formed exclusively by conformal 
symmetries. We call {\em proper causal symmetries} the causal symmetries 
which are not conformal transformations.

The set $\C(V)$ is invariant against conformal rescaling, that is 
$\C(V,e^{\sigma}\G)=\C(V,\G)$ for all differentiable
functions $\sigma$, so the assertion that $\f$ is a causal symmetry is
a conformally invariant one. Moreover, if $(V,\G_{1})$ and
$(V,\G_{2})$ are isocausal ---meaning that there are mutual causal relations
$\phi$ and $\psi$, see PII---, then there is a one-to-one mapping between 
$\C(V,\G_{1})$ and $\C(V,\G_{2})$ because if $\f\in \C(V,\G_{1})$ 
then one can easily construct a causal symmetry
of $(V,\G_{2})$ (say $\phi\circ\f\circ\psi$), and vice versa. These two facts
allow us to claim that causal symmetries keep the causal structure
---in the sense of PII--- invariant.

For any non-zero rank-$r$ future tensor ${\bf T}\in\DP^+_r(V)$ we define the
set of its {\em principal null directions}, denoted by $\mu(\T)$,
as the set of future-directed vectors $\k$ such that $\T(\k,\dots,\k)=0$.
This immediately implies that $\k$, being causal, must in fact be null,
(property 2.3 in PI). This concept is a close relative of the
one presented in \cite{PP}, which itself is a generalization of
the principal null directions of the Weyl tensor. 
By definition, the set of canonical null directions (section 4 of PII) 
of a causal symmetry $\f$ is simply $\mu(\f^{*}\G)$, whose elements are the null
eigendirections of $\f^{*}\G$. Then we have
\be
\hspace{-15mm}
\f\in\C(V) \Longrightarrow \f'[\mu(\f^{*}\T)]\subseteq \mu(\T) \,\,
\mbox{and}\,\, \mu(\f^{*}\T)\subseteq \mu(\f^{*}\G), \, \forall\T\in\DP^+_r(V).
\label{pndT}
\ee
Recall that if $\f\in\C(V)$, then $\f^{*}\T\in\DP^+_r(V)$ for all
$\T\in\DP^+_r(V)$ (proposition 3.1 of PII). The first assertion follows
immediately from $(\f^{*}\T)(\k,\dots,\k)=\T(\f^{'}\k,\dots,\f^{'}\k)$,
and the second from the fact that $\f'\k$ is null if $\k\in\mu(\f^{*}\T)$
---using again property 2.3 in PI---, so that
$0=\G(\f'\k,\f'\k)=(\f^{*}\G)(\k,\k)$. Important corollaries of (\ref{pndT})
are (i) $\mu(\f^{*}\G)=\emptyset \ \Longrightarrow \
\mu(\f^{*}\T)=\emptyset$ and (ii) $\mu(\T)=\emptyset \ \Longrightarrow \
\mu(\f^{*}\T)=\emptyset , \ \forall \T\in\DP^+_r(V)$.

Of course, the $\mu$-sets depend on the point 
of the manifold. However, using the techniques of algebraic 
decompositions of spacetimes \cite{HALL,RAUL} one can see 
that $V$ splits in open subsets where $\mu(\f^{*}\G)$ has a constant 
number of linearly independent elements. Henceforth, we will 
work on one of these subsets and assume that $\G$ is analytic 
there.

As usual with general symmetries, we are interested in the possibility
of constructing one-parameter groups of causal symmetries, and their
infinitesimal versions. Let $\{\f_{s}\}_{s\in I}$ be a local
one-parameter group of transformations where $I\subseteq \r$ is an
open interval and $s$ its canonical parameter. 
When do these groups contain elements of $\C(V)$? 
A first answer comes from the following fact: if
$\{\f_s\}_{s\in[0,\epsilon)}\subset \C(V)$ with $[0,\epsilon)\subset I$,
then every element of $\{\f_{s}\}_{s\in \r^{+}\cap I}$ is a causal
symmetry. This follows because every $s_0\in \r^+$ can be written as a
finite sum of numbers $s_1,\dots ,s_j\in [0,\epsilon)$,
so that $\f_{s_0}=\f_{s_1+\dots+s_j}=\f_{s_1}\circ\dots\circ\f_{s_j}$
is a composition of causal symmetries and thus a causal symmetry itself.

Now suppose that under the above hypotheses $\f_{s_0}$ is a conformal 
transformation for $|s_0|\in I\cap\r^+$ 
and let $\k$ be an arbitrary future-directed null 
vector. Since a conformal transformation maps null vectors onto null 
vectors, $\f^{'}_{|s_0|}\k$ is also null. Then
$\f^{'}_s\k$ is null for all $s\in [0,|s_0|]$ because
$\f^{'}_{|s_0|}\k=\f^{'}_{s_1}[\f^{'}_{s_2}\k]$
with $|s_0|=s_1+s_2$ where $s_1,s_2\in (0,|s_0|)$, and using that
$\f_{s_{1}},\f_{s_{2}}$ are causal symmetries one has that
$\f^{'}_{s_2}\k$ is causal and then (proposition 3.2
of PII) it must necessarily be null, proving that
$\f_s$ are conformal transformations $\forall s\in (-|s_{0}|,|s_0|)$ 
since they map null vectors onto null vectors 
(theorem 4.2 of PII). In turn, this implies that $\{\f_s\}_{s\in I}$ 
consists of conformal symmetries due to the group property of such 
transformations. We summarize this in the next result.
\begin{res}
Suppose $\{\f_s\}_{s\in I}$ is a local one-parameter
group of transformations such that $\{\f_s\}_{s\in I\cap\r^+}\subset\C(V)$.
Then either $\f_s$ is a conformal transformation for every value of $s\in I$
or $\{\f_s\}_{s\in I}$ contains no conformal transformations other than the
identity ($=\f_{0}$).
\label{no-conf}
\end{res}

\vspace{-7mm}
An immediate corollary of this result is that there cannot be
{\em cyclic} submonoids of proper causal symmetries, so that the
orbits of these submonoids can never be closed. For if $\{\f_s\}_{s\in
S^1}$ were an effective realization of the circle formed by causal
symmetries, then with the usual parameterization $\f_{2\pi}$ would be the
identity map, that is to say, a (conformal) isometry, so that the
whole subgroup would be conformal. Obviously, we will be interested 
in cases {\em with} proper causal symmetries. 
The set $\{\f_s\}_{s\in I\cap\r^+}$ will be called a
{\em local one-parameter submonoid of causal symmetries}
if it is a subset of $\C(V)$.

Our first fundamental result regarding these submonoids 
is that for $s>0$ the sets $\mu(\f^{*}_{s}\G)$ are independent of $s$,
and their elements are simply the null vector fields which remain null under
the action of $\{\f_s\}_{s\in I}$. To prove this,
let $\k$ be a null future-directed vector in $\mu(\f^*_{s_0}\G)$ for 
$s_0\in I\cap\r^+$, which is equivalent to $\f^{'}_{s_0}\k$ being null 
and future-directed. Then, reasoning in much the same way as we did 
in Result \ref{no-conf}, we get that $\f^{'}_s\k$ is null future-directed 
$\forall s\in [0,s_0]$ which is only possible (proposition 4.1 of PII) if 
$\k\in\mu(\f^{*}_s\G)$. Then, the analytic function 
$f_{\k}(s)\equiv (\f^{*}_s\G)(\k,\k)=\G(\f^{'}_s\k,\f^{'}_s\k)$ 
vanishes on the open interval $(0,s_0)$ and hence it must vanish on 
the whole $I$. As a bonus we also deduce that $\f^{'}_s\k$ is null for 
all $s\in I$. Let $\xiv$ be the infinitesimal generator of $\{\f_s\}_{s\in I}$. 
We denote simply by 
$\mu_{\xiv}$ the set $\mu(\f^{*}_{s}\G)$ for {\em any} $s>0$
and their elements are called the {\em canonical null directions} of the
submonoid of causal symmetries. All the elements of $\mu_{\xiv}$ are 
eigenvectors of $\f^{*}_s\G$
with the {\em same} eigenvalue $\lambda_s$ for each $s$. From the above
$\f^{'}_{s}(\mu_{\xiv})=\mu_{\xiv}$ for every $s\in I$, which allows 
to get the following fundamental property.
\begin{res}
If $\{\f_s\}_{s\in I\cap\r^+}\subset\C(V)$, then for every $\T\in\DP^+_r(V)$,
$\f'_{s}[\mu(\f^{*}_{s}\T)]=\mu(\T)\cap\mu_{\xiv}$.
\label{TENSOR-NULL}
\end{res}

\vspace{-7mm}
The inclusion $\f^{'}_s(\mu(\f^{*}_s\T))\subseteq\mu_{\xiv}\cap\mu(\T)$
follows directly from (\ref{pndT}) if we take into account that
$\f^{'}_{s}(\mu_{\xiv})=\mu_{\xiv}$, $\forall s\in I$. Conversely, pick up
any $\k\in\mu_{\xiv}\cap\mu(\T)$ so that
$0=\T(\k,\dots,\k)=(\f^{*}_s\T)(\f^{'}_{-s}\k,\dots,\f^{'}_{-s}\k)$
for $s>0$. As $\k\in\mu_{\xiv}$,
$\f^{'}_{-s}\k$ must be null for every $s\in I$, and since
$\f^{*}_s\T\in\DP^+_r(V)$ we get that
$\f^{'}_{-s}\k\in\mu(\f^{*}_s\T)$ from what 
$\mu_{\xiv}\cap\mu(\T)\subseteq\f^{'}_s(\mu(\f^{*}_s\T))$ follows.
Result \ref{TENSOR-NULL} implies that if
$\mu_{\xiv}\cap\mu(\T)=\emptyset$ then $\f^{*}_s\T$ has no
principal null directions for every $s>0$, while if
$\mu_{\xiv}\subseteq \mu(\T)$ then $\mu(\f^{*}_s\T)=\mu_{\xiv}$.

As $\mu_{\xiv}$ is a set of null directions, it is not a vector space.
Nevertheless, we can pick up a maximum number of {\em linearly independent}
null vector fields $\{\k_1,\dots,\k_m\}$ belonging to $\mu_{\xiv}$, so
that $Span\{\mu_{\xiv}\}$ is invariant under the linear transformations
$\f^{'}_s$, being the eigenspace associated to $\lambda_s$ for $s>0$.
The number $m\equiv$dim($Span\{\mu_{\xiv}\}$) is intrinsic to the
submonoid of causal symmetries.  
Let $\O=\K_1\wedge\dots\wedge\K_m$ be a characteristic $m$-form over 
$Span\{\mu_{\xiv}\}$, where $\K_1,\dots,\K_m$ are the one-forms 
associated to $\k_1,\dots,\k_m$. From the previous results it is easy 
to see that\footnote{In the cases $m=1,2$ we can
further establish the property $\f^{'}_s\k\propto\k$,
$\forall\k\in\mu_{\xiv}$, as is obvious.}
\be
\f^{*}_s\O=\sigma_s\O, \hspace{3mm} \mbox{for some} \hspace{1mm}
\sigma_{s}\in C^{\infty}(V), \,\, \forall s\in I \hspace{6mm}
\Longleftrightarrow \hspace{6mm} \lie\O=\g\O \label{Omega2}. 
\ee

The set $\mu_{\xiv}$ plays a key role in the study of the causal 
symmetries. Furthermore, it allows to set up a convenient
classification of causal (and more general)
symmetries, according to the number $m$ defined above. 
When $m=n$ we recover the conformal
symmetries, while for $1\leq m<n$, we can speak of 
{\em $\fr{m}{n}$-partly conformal}
symmetries, as they leave invariant the $m$ independent null directions
within $Span\{\mu_{\xiv}\}$. This view will be further supported later
by the equations of the infinitesimal causal symmetries.
Thus, we have a classification of causal symmetries,
which split up into $n+1$ different types according to whether
$m=0,\ldots ,n$. This is a more justified and better defined algebraic
classification than the one recently outlined
in \cite{sergi}. It also includes, for $m=1$, the newly studied
case of Kerr-Schild vector fields \cite{KERR-SCHILD}. It is worth noting that
the symmetries closer to conformal ones are those with
$m=n-1$, rendering those with $m=1$ ---in particular those of
\cite{KERR-SCHILD}-- as the ``less conformal'' among the partly conformal
symmetries. We  will also see that the case with $m=1$ is degenerate 
within this classification.

We have to know how to compute $\mu_{\xiv}$ or the 
generalization of the conformal property $\lie\G\propto\G$ to the causal 
symmetries. To that end, we need a lemma.
\begin{lem}
Let $\{\T_s\}$ be a one-parameter family, differentiable in $s$,
of rank-$r$ (covariant) tensors such that $\T_{s_{0}}=0$ for some fixed $s_0$. 
Assume that $\T_s\in\DP^+_r(V)$ for all $s\in [s_0,s_0+\epsilon)$. Then 
$d\T_{s}/ds|_{s=s_{0}}\equiv \dot{\T}_{s_0}\in\DP^+_r(V)$ 
(or its contravariant counterpart). 
\label{basic-lem}
\end{lem}
To prove it, define functions $f_{\u_1,\dots,\u_r}(s)\equiv\T_s(\u_1,\dots,\u_r)$ 
where $\u_1,\dots,\u_r$ are any future-directed causal vectors. 
Clearly $f_{\u_1,\dots,\u_r}(s_{0})=0$ while 
$f_{\u_1,\dots,\u_r}(s)\geq 0$ for $s\in [s_0,s_0+\epsilon)$, which 
immediately implies $0\leq df_{\u_1,\dots,\u_r}/ds|_{s_{0}}=
\dot{\T}_{s_0}(\u_1,\dots,\u_r)$.
As a first application, we are now ready to get the sought expression of $\lie\G$.
\begin{res} There exists a smooth function $\alpha$ such that
$(\lie\G-\alpha\G)\in\DP^{+}_2(V)$.\label{basic}
\end{res}
Indeed, $\f_{s}^{*}\G\in\DP^{+}_2(V)$ for every $s\in\r^+\cap I$,
hence we can apply the canonical {\em decomposition theorem}
(theorem 4.1 of PI) to such causal tensors to get 
\be
\f^{*}_s\G=\sum_{p=m}^n{\bf T}\{\O_{[p]}(s)\}=
\sum_{p=m}^{n-1}{\bf T}\{\O_{[p]}(s)\}+A^2_{s}\, \G,
\label{DECOMP}
\ee
where $A_{s}$ is a differentiable function such that $A_{0}=1$,
and ${\bf T}\{\O_{[p]}(s)\}$ are the superenergy tensors
\cite{SUP,PI} of adequate simple $p$-forms $\O_{[p]}(s)$. 
The general formula for the superenergy tensor of a $p$-form ${\bf \Sigma}$ 
is \cite{SUP,PI}:
\be
T\{{\bf \Sigma}\}_{ab}=\fr{(-1)^{p-1}}{(p-1)!}
\left(\Sigma_{ac_2\dots c_p}\Sigma_{b}^{\ c_2\dots c_p}-\fr{1}{2p}\mbox{g}_{ab}
\Sigma_{c_1\dots c_p}\Sigma^{c_1\dots c_p}\right).
\label{s-simple}
\ee 
Each term appearing in equation (\ref{DECOMP}) is in
$\DP^+_2(V)$ and we have distinguished
the extreme value $p=n$ because the corresponding tensor is
proportional to the metric (PI). Therefore,
the family $\T_{s}=\f^*_{s}\G -A^2_s\G$ 
satisfies the conditions of Lemma \ref{basic-lem} with $s_{0}=0$
from what Result \ref{basic} follows with $\alpha\equiv 
dA^2_s/ds|_{s=0}$ by using that $\lie\G =d(\f^{*}_s\G)/ds|_{s=0}$.

We can apply now the decomposition theorem 4.1 of PI to the future tensor
$\lie\G-\alpha\G$. To do it, we must know the
set $\mu(\lie\G-\alpha\G)$. 
As $\mu(\f^{*}_{s}\G)=\mu_{\xiv}$, $\f^{*}_s\G$
always have the null vector fields of $\mu_{\xiv}$ as
eigendirections so that we can consistently choose in (\ref{DECOMP})
$\O_{[m]}(s)\propto \O$ for all $s\in I\cap \r^+$ if $m>0$. Thus, we 
will use the notation $\S\equiv\T\{\O\}$ from now on. From the results in PI, 
$\S^2$ is proportional to $\G$ so that we will also assume that $\S$ has
been normalized if $m>1$, that is, $S_{ac}S^c{}_{b}=\mbox{g}_{ab}$. The case
$m=1$ is degenerate in the sense that $S_{ac}S^c{}_{b}=0$, equivalent to
$\S =\K \otimes \K$ where $\K$ is a representative of the unique
canonical null direction. It is quite simple to deduce that the elements
of $\mu_{\xiv}$ are among the null eigenvectors of $\lie\G$ by using 
that, for any $\k\in\mu_{\xiv}$,
$\f^{*}_{s}\G(\cdot ,\k)=\lambda_s\G(\cdot ,\k)$, $\forall s\in I$.
This implies that $\mu_{\xiv}\subseteq \mu(\lie\G-\alpha\G)$.
Now, assume that there were
a $\k\in\mu(\lie\G-\alpha\G)\setminus \mu_{\xiv}$. Then $\f'_{s}\k$ would be
timelike for $s>0$ so that, using e.g. Lemma 2.5 in PI, we could write
$\f'_{s}\k=c_{s}\k +\vec{n}_{s}$ where the $\vec{n}_{s}$ are null and 
future directed and $c_{s}>0$ such that
$\vec{n}_0=\vec{0}$, $c_0=1$. But then the family $\f'_{s}\k-c_{s}\k$ 
would satisfy the hypotheses of Lemma \ref{basic-lem} with $s_{0}=0$,
proving that there would be a function $c$ such that $-\lie\k+c\k$ is
future pointing. On the other hand, using that
$\k\in\mu(\lie\G-\alpha\G)$ we get $0=\lie[\G(\k,\k)]=2\G(\lie\k,\k)$
so that $-\lie\k+c\k$ and $\k$, being both future pointing
and orthogonal to each other, would necessarily be null and proportional,
leading to $\lie\k\propto \k \Longleftrightarrow \f'_{s}\k\propto\k$, 
which would mean $\k\in\mu_{\xiv}$ in contradiction. Thus,
$\mu_{\xiv}=\mu(\lie\G-\alpha\G)$ and we have
\be
\lie\G=\alpha\G+\beta\S+\Q
\label{1ST}
\ee
where $\Q$ is a symmetric rank-2 future tensor such that
$\mu(\Q)\supset \mu_{\xiv}$ whence (see PI)
$$
Q_{a}{}^b\Omega_{ba_{2}\dots a_{m}}=\lambda \Omega_{aa_{2}\dots a_{m}},
\hspace{5mm} Q_{a}{}^c(\mbox{g}_{cb}+S_{cb})=\lambda (\mbox{g}_{ab}+S_{ab})
\Longrightarrow\hspace{3mm}Q_{a}{}^cS_{cb}=Q_{b}{}^cS_{ca}
$$
and $\beta>0$, $\lambda\geq 0$ are smooth functions. The first equation 
comes 
from $Q_{a}{}^bk_b=\lambda k_a$ $\forall \k\in\mu_{\xiv}$, while 
the second follows because $\G+\S$ is the projector onto 
$Span\{\mu_{\xiv}\}$.

Relations (\ref{Omega2}) and (\ref{1ST}) are the fundamental equations
of this letter. They are ``stable'' under
repeated application of $\lie$, that is to say, the structure of their
right hand sides remains the same. This is clear for (\ref{Omega2}). To
prove it for (\ref{1ST}) we need to know the Lie derivatives of 
tensors of the type of $\S$ or $\Q$. For $\S$ this can be easily done
by using its explicit expression $\S=\T\{\O\}$ (eq. (\ref{s-simple})) which meets
the normalization requirements if we put
$\Omega_{c_1\dots c_m}\Omega^{c_1\dots c_m}
=2m!(-1)^{m-1}$ when $m>1$. Then, by means of (\ref{Omega2}) and (\ref{1ST}) 
we readily arrive at
\be
\lie S_{ab}=\alpha S_{ab}+\beta \mbox{g}_{ab}+Q_{ac}S^{c}_{\ b},
\hspace{1cm} (m>1).\label{2ND}
\ee
As is clear from their derivation, eqs.(\ref{Omega2}), (\ref{1ST})
and (\ref{2ND}) are not independent and, actually,
(\ref{Omega2},\ref{1ST}) are equivalent to (\ref{1ST},\ref{2ND})
where, due to the chosen normalization, one necessarily has
$2\gamma =m(\alpha +\beta +\lambda )$ for $m>1$. In the degenerate case 
$m=1$, $\alpha,\beta$ and $\gamma$ can be kept arbitrary and 
the equation replacing (\ref{2ND}) is just 
$\lie\S=2\gamma\S\Longleftrightarrow \lie\K =\gamma\K$ ($m=1$).

With regard to tensors of type $\Q$, we need an intermediate result which asserts
that for two given future tensors $\T_1$ and $\T_2$ with $\mu(\T_1)=\mu(\T_2)$ 
and $\mbox{dim}(Span\{\mu(\T_1)\})\geq 1$ we can always find a positive 
$\alpha_{12}$ and a future tensor $\R_{1}$ such that $\T_1=\alpha_{12}\T_2+\R_1$.
The proof is rather straightforward by noticing the 
existence of an orthonormal basis which diagonalizes both $\T_1$ and $\T_2$. 
Thus, since $\mu(\f^{*}_{s}\Q)=\mu_{\xiv}$ (use Result \ref{TENSOR-NULL}), we can 
apply this to the causal tensors $\f^*_{s_1}\Q$ and $\f^{*}_{s_2}\Q$ for 
$s_1,s_2\in[0,\epsilon)$ and write 
$$
\f^{*}_{s_2}\Q=\alpha_{s_2,s_1}\f^{*}_{s_1}\Q+\R_{s_2,s_1},\ \ \
\R_{s_2,s_1}\in\DP^+_2(V)
$$
where we can choose $\alpha_{s_1,s_1}=1$ and $\R_{s_{1},s_{1}}=0$. Applying Lemma 
\ref{basic-lem} to the family 
$\f^{*}_{s}\Q-\alpha_{s,s_1}\f^{*}_{s_1}\Q$ with $s_{1}$ fixed we 
get
$$
\left(\left.\fr{d(\f^{*}_{s}\Q)}{ds}\right|_{s=s_1}-
\left.\fr{d\alpha_{s,s_1}}{ds}\right|_{s=s_1}(\f^{*}_{s_1}\Q)\right)
\in\DP^+_2(V),\ \ s_1\in [0,\epsilon)
$$
from what, by putting $s_1=0$, the desired result follows:
\begin{res}
For every $\Q\in \DP^+_{2}(V)$ with 
$\mu (\Q)\supseteq \mu_{\xiv}\neq \emptyset$ 
there is a smooth function $\psi$ such that 
$\lie\Q-\psi\Q\in\DP^+_2(V)$ and $\mu (\lie\Q-\psi\Q)\supseteq \mu_{\xiv}$.
\end{res}
In particular, we can apply this result to $\lie\G-\alpha\G$ to get 
as a corollary the existence of functions $\alpha_1,\dots,\alpha_r,\ldots$ 
for all natural $r\in \NAT$ such that
$$
(\lie-\alpha_r)\cdots (\lie-\alpha_1)(\lie-\alpha)\G\in\DP^+_2(V), 
\,\,\, \forall r\in\NAT
$$ 
where at any level $r$ the set of principal null directions always 
includes $\mu_{\xiv}$. This is the required property on the stability of 
(\ref{1ST}).

Two remarkable equations deducible from (\ref{1ST}) and (\ref{2ND}) are ($m>1$)
$$
\lie S^a{}_{b} =0, \hspace{1cm} \lie (\G +\S) =(\alpha +\beta
+\lambda )(\G +\S).
$$
These formulae support the claim that causal symmetries define partly
conformal Killing vectors, being conformal on $Span\{\mu_{\xiv}\}$.
As, on the other hand,
$\lie (\mbox{g}_{ab} -S_{ab})=(\alpha -\beta)(\mbox{g}_{ab} -S_{ab})+Q_{a}{}^c(\mbox{g}_{cb} -S_{cb})$
they will also be conformal on the orthogonal subspace $\perp
Span(\mu_{\xiv})$ if and only if 
$Q_{a}{}^c(\mbox{g}_{cb} -S_{cb})\propto (\mbox{g}_{ab}
-S_{ab})$, which is only possible if $\Q\propto \G$. This is
equivalent, by redefining $\alpha$, to $Q_{ab}=0$ (and hence $\lambda=0$).
Therefore we say that a causal symmetry is {\em pure} if $\Q=0$.
The general case with $\Q$ non-vanishing will be dealt with in 
\cite{S}. 
Observe that the cases $m=n-1,n$ are always pure.
The generating vector fields of pure causal symmetries
satisfy then ($m\neq 1$)
\be
\lie \mbox{g}_{ab}=\alpha \mbox{g}_{ab}+\beta S_{ab}, \hspace{1cm}
\lie S_{ab}=\alpha S_{ab}+\beta \mbox{g}_{ab}.\label{2STP}
\ee
In the degenerate situation $m=1$, the pure case can also be defined
by the vanishing of $\Q$ and the corresponding equations are
\be
\lie \mbox{g}_{ab}=\alpha \mbox{g}_{ab}+\beta k_ak_b,\hspace{1cm} \lie k_a=\gamma k_a
\label{DEGP}
\ee
which include ($\alpha =0$) the Kerr-Schild vector fields
of \cite{KERR-SCHILD}. Of course, $\alpha$ and $\beta$ (and $\gamma$
if $m=1$) actually depend on $\xiv$, so
they will be called the gauge functions as in \cite{KERR-SCHILD}. In fact
eqs.(\ref{2STP}) (or (\ref{DEGP})) are also sufficient, even if 
$\S$ is just a future tensor or if $\K$ is just causal:
\begin{res}
A vector field $\xiv$ which satisfies (\ref{2STP}) (respectively (\ref{DEGP}))
with $\beta \S\in\DP^+_{2}(V)$ and {\rm dim($Span\{\mu (\S)\}\neq 1$)}
(resp.\ $\S=\K\otimes\K$ with causal 
$\K$) generates a one-parameter submonoid of causal symmetries 
$\{\f_s\}_{s\in I\cap\r^+}$ with $\mu_{\xiv}=\mu(\beta \S)$. 
\end{res}
To prove this when $m\neq 1$, we use
the general formula $\f^{*}_{s}(\lie\T)=d(\f^{*}_{s}\T)/ds$ which by
integration immediately leads to
$\f^{*}_{s}(\G +\S)=\exp\{\int^s_{0}\f^{*}_{t}(\alpha +\beta)dt\}(\G +\S)$
and $\f^{*}_{s}(\G -\S)=\exp\{\int^s_{0}\f^{*}_{t}(\alpha -\beta)dt\}(\G
-\S)$, from where we deduce
$$
\f^{*}_{s}\G =\exp\left\{\int^s_{0}\alpha (\f_{t})dt\right\}\left[\cosh
\left(\int^s_{0}\beta (\f_{t})dt\right)\G +
\sinh \left(\int^s_{0}\beta (\f_{t})dt\right)\S\right]
$$
which are clearly future tensors for all $s>0$ if 
$\beta\S\in\DP^+_{2}(V)$, so
that $\{\f_{s}\}_{s\in \r^+\cap I}$ is a submonoid of causal symmetries.
The proof for the other case ($m=1$) is analogous.

Observe that $\beta$ must have a definite sign, implying that the 
vector fields satisfying (\ref{2STP}) do not form a vector space, but 
only a {\em wedge} or {\em cone}, see \cite{SEMIGROUP},
of a vector space. Nevertheless, the study of (\ref{2STP}) and
(\ref{DEGP}) has an interest on its own right, independently of the  
gauges signs, as they always define pure partly conformal symmetries
(albeit possibly not causal) 
with a vector space structure. Its general study will
be addressed elsewhere \cite{S}, but in the rest of the letter we give 
some preliminary results. First of all, (\ref{2STP}) defines a Lie-algebra 
structure: if
$\xiv_1$ and $\xiv_2$ comply with eqs.(\ref{2STP}) with gauges
$\alpha_{\xiv_1}$, $\beta_{\xiv_1}$
and $\alpha_{\xiv_2}$, $\beta_{\xiv_2}$ respectively then their Lie
bracket $[\xiv_2,\xiv_1]$ also satisfies (\ref{2STP}) with
gauges
\[
\alpha_{[\xiv_2,\xiv_1]}=\pounds_{\xiv_2}\alpha_{\xiv_1}-
\pounds_{\xiv_1}\alpha_{\xiv_2},\ \ \beta_{[\xiv_2,\xiv_1]}=
\pounds_{\xiv_2}\beta_{\xiv_1}-\pounds_{\xiv_1}\beta_{\xiv_2}.
%\label{GAUGE}
\]
A similar computation leads the same conclusion for the degenerate
case $m=1$\addtocounter{footnote}{+4}\footnote{Actually, this reasoning is 
independent of the properties of $\S$ (or $\K$), so that (\ref{2STP}) 
(or (\ref{DEGP})) define Lie algebras for {\em any} tensor field $\S$ (or 
{\em any} one-form $\K$).}. 
These Lie algebras define the corresponding transformations
groups whose generators satisfy
(\ref{2STP}) (or (\ref{DEGP})) and they can be, in
certain cases, infinite dimensional. These groups will
contain submonoids of causal symmetries only when the gauges $\beta_{\xiv}$
have a sign. Thus, if $\xiv_1$ and $\xiv_2$ generate pure causal
symmetries with $\mu_{\xiv_{1}}=\mu_{\xiv_{2}}$ then $\pm[\xiv_2,\xiv_1]$ 
will also be such a kind of generator only if 
$\beta_{[\xiv_2,\xiv_1]}=\pounds_{\xiv_2}\beta_{\xiv_1}-
\pounds_{\xiv_1}\beta_{\xiv_2}$ does not vanish anywhere.

An example of physical relevance is provided by the so-called warped 
products, that is to say, Lorentzian manifolds 
of the form $V_{1} \times \hat{V}$ with metrics of type 
$\G=\G^1-R^2 \hat{\G}$ where $\G^{1},\hat{\G}$ are metrics on $V_{1},\hat{V}$ 
respectively, and $R$ is a non-vanishing function on $V_{1}$. 
Here we concentrate on the case where $(V_{1},\G^{1})$ is $m$-dimensional and 
Lorentzian so that 
$$
ds^2=\mbox{g}^1_{\alpha\beta}(x^\g)dx^{\alpha}dx^{\beta}-R^2(x^\g)d\S^2_{n-m},\ \ 
d\S^2_{n-m}=\hat{\mbox{g}}_{AB}(x^{C})dx^Adx^B
$$ 
where $\{x^\g\}$ ($\alpha,\beta,\g =0,\ldots,m-1$) are coordinates on 
$V_{1}$ and $d\S^2_{n-m}$ is the 
positive-definite line-element of $(\hat{V},\hat{\G})$ whose 
coordinates are $\{x^{A}\}$ ($A,B,C =m,\ldots,n-1$). We seek the pure causal
symmetries with $\O=\rho\ dx^0\wedge\dots\wedge dx^{m-1}$ where
$\rho=\sqrt{2\ \mbox{det}(\G^1)}$ to meet the needed normalization. 
Equations (\ref{2STP}) imply that $\xiv$ decomposes as $\xiv=\xiv_1+\xiv_2$ with 
$\xiv_1=\xi^{\alpha}(x^\g)\d_{\alpha}$, $\xiv_2=\xi^A(x^B)\d_A$, and 
also
$$
\pounds_{\xiv_2}\hat{\G}=\left(\alpha-\beta- 
R^{-2}\xiv_1(R^2)\right)\hat{\G},\ \ 
\pounds_{\xiv_1}\G^1=(\alpha+\beta)\G^1 \ \Longrightarrow \
\fr{1}{\rho}\pounds_{\xiv_1}\rho+\d_{\g}\xi_{1}^\g =\fr{1}{2}m(\alpha+\beta).
$$
Notice that $\xiv_1$ and $\xiv_2$ are conformal 
symmetries of $\G^1$ and $\hat{\G}$ respectively. The number of 
independent $\xiv$ depends on $n,m$ and the particular $V_{1},\hat{V}$, 
and it can be finite ($n-m,m>2$) or infinite (in some cases with 
$n-m\leq 2$ or $m=1,2$).

Simple examples of the above are provided by $n$-dimensional Minkowski spacetime 
in Cartesian coordinates. Its pure causal symmetries
with $\perp\!\! Span(\mu_{\xiv})=\left<\partial_{x^{n-1}}\right>$ 
($m=n-1$) are given by
$\xiv =\xiv_1 +F(x^{n-1})\partial_{x^{n-1}}$
where $\xiv_1$ is any conformal Killing vector of the $(n-1)$-dimensional
Minkowski spacetime and $F$ is arbitrary. Thus,
in this case $\xiv$ depends on $n(n+1)/2$ parameters and
a function of one coordinate. Finite-dimensional cases
also appear, as of course the strictly conformal case with $m=n$. 
Another example arises by taking (say) $n=6$ and
$Span\{\mu_{\xiv}\}=\left<\partial_{x^0},\partial_{x^1},\partial_{x^2}\right>$,
so that $\xiv_1$ and $\xiv_2$
are conformal Killing vectors of the 3-spaces $Span\{\mu_{\xiv}\}$
and $\perp\!\! Span\{\mu_{\xiv}\}$, respectively. Hence, now 
the general
$\xiv$ depends on $10+10=20$ arbitrary parameters.

A more interesting situation comprises
the spherically symmetric spacetimes, in which $(\hat{V},\hat{\G})$ has  
positive constant curvature and $(V_1,\G^1)$ is 2-dimensional so that
$ds^2_{1}=2e^{f(u,v)}dudv$, $\O =\sqrt{2}e^{f}du\wedge dv$ and
$\mu_{\xiv}=\{\partial_{u},\partial_{v}\}$. This has a clear physical 
interpretation for $\mu_{\xiv}$ are the {\em radial} null directions. 
The previous calculation particularizes now to 
$\xiv_1=\xi^u(u)\d_{u}+\xi^v(v)\d_{v}$ with
$\xiv_{1}(f)+\xi^{u}_{,u}+\xi^v_{,v}=\alpha+\beta$.
Observe that the gauges are determined by the data $f(u,v),R(u,v)$ and
the particular $\xiv_{2}$ and $\xiv_1$.
Thus the general $\xiv$ depend on two arbitrary 
functions $\xi^{u}(u),\xi^{v}(v)$, plus 
the number of independent conformal Killing vectors $\xiv_{2}$ of 
the $(n-2)$-sphere. For instance, if $n=4$ this conformal 
group has 6 independent parameters and is isomorphic 
to the Lorentz group.

Several other symmetries already appeared in the literature are also included in
the causal symmetries, such as the conformal Killing
vectors, the Kerr-Schild vector fields \cite{KERR-SCHILD}, some
examples given in \cite{LOW}, or the transformations studied in
\cite{Bon,Mars,sergi}.

\section*{Acknowledgements}
We thank L. Bel, G. Bergqvist, B. Coll, M. Mars and R. Vera for a careful reading
of the manuscript and the improvements they suggested. This work has been carried
out under grant 9/UPV00172.310-14456/2002 of the University of the 
Basque Country.


\section*{References}
\begin{thebibliography}{99}
\bibitem{PI} Bergqvist G and Senovilla J M M  2001 {\it Class. Quantum Grav.}
{\bf 18} 5299
\bibitem{Bon} Bonanos S 1992 {\it Class. Quantum Grav.} {\bf 9} 697
\bibitem{KERR-SCHILD} Coll B, Hildebrandt S R and Senovilla J M M 2001
{\it Gen. Rel. Grav} {\bf 33} 649
\bibitem{PII} Garc\'{\i}a-Parrado A and Senovilla J M M 2003 
{\it Class. Quantum Grav.} {\bf 20} 625 
\bibitem{S} Garc\'{\i}a-Parrado A and Senovilla J M M, in preparation
\bibitem{HALL} Hall G S 1996 {\it Class. Quantum Grav.} {\bf 13} 1479
\bibitem{LOW} Harris S G and  Low R J 2001 {\it Class. Quantum Grav.} {\bf 18} 27
\bibitem{sergi} Hildebrandt  S R 2002 {\it Gen. Rel. Grav.} {\bf 34} 65
\bibitem{SEMIGROUP} Hilgert J, Hofmann K H  and Lawson J D 1989
{\it Lie groups, Convex Cones and Semigroups.} (Oxford Science Publications)
\bibitem{Mars} Mars M 2001 {\it Class. Quantum Grav.} {\bf 18} 719
\bibitem{PP}  Pozo J M and Parra J M 2002 {\it Class. Quantum Grav.}
{\bf 19} 967
\bibitem{SUP} Senovilla J M M 2000 {\it Class. Quantum Grav.} {\bf 17} 2799
\bibitem{RAUL} Senovilla J M M and Vera R 1999 {\it Class. Quantum Grav.} {\bf 16} 1185
\end{thebibliography}
\end{document}